\documentclass[sigconf]{aamas}

\usepackage{balance} 

\usepackage{algorithm}

\usepackage{subfiles}
\usepackage{subcaption}
\usepackage{algorithmicx}
\usepackage{algpseudocode}
\usepackage{ragged2e}
\usepackage{graphicx}

\setcopyright{none}
\acmConference[AAMAS '25]{Proc.\@ of the 24th International Conference
on Autonomous Agents and Multiagent Systems (AAMAS 2025)}{May 19 -- 23, 2025}
{Detroit, Michigan, USA}{A.~El~Fallah~Seghrouchni, Y.~Vorobeychik, S.~Das, A.~Nowe (eds.)}
\copyrightyear{2025}
\acmYear{2025}
\acmDOI{}
\acmPrice{}
\acmISBN{}

\title[Enhancing Heterogeneous Multi-Agent Cooperation in Decentralized MARL via GNN-driven Intrinsic Rewards]{Enhancing Heterogeneous Multi-Agent Cooperation in Decentralized MARL via GNN-driven Intrinsic Rewards}

\author{Jahir Sadik Monon}
\affiliation{
  \institution{Independent University Bangladesh}
  \city{Dhaka}
  \country{Bangladesh}
  }

\author{Deeparghya Dutta Barua}
\affiliation{
  \institution{Penta Global Ltd}
  \city{Dhaka}
  \country{Bangladesh}
  }
\author{Md Mosaddek Khan}
\affiliation{
  \institution{University of Dhaka, Bangladesh}
  \city{Dhaka}
  \country{Bangladesh}
  }

\begin{abstract}
Multi-agent Reinforcement Learning (MARL) is emerging as a key framework for various sequential decision-making and control tasks. Unlike their single-agent counterparts, multi-agent systems necessitate successful cooperation among the agents. The deployment of these systems in real-world scenarios often requires decentralized training, a diverse set of agents, and learning from infrequent environmental reward signals. These challenges become more pronounced under partial observability and the lack of prior knowledge about agent heterogeneity. While notable studies use intrinsic motivation (IM) to address reward sparsity or cooperation in decentralized settings, those dealing with heterogeneity typically assume centralized training, parameter sharing, and agent indexing. To overcome these limitations, we propose the CoHet algorithm, which utilizes a novel Graph Neural Network (GNN) based intrinsic motivation to facilitate the learning of heterogeneous agent policies in decentralized settings, under the challenges of partial observability and reward sparsity. Evaluation of CoHet in the Multi-agent Particle Environment (MPE) and Vectorized Multi-Agent Simulator (VMAS) benchmarks demonstrates superior performance compared to the state-of-the-art in a range of cooperative multi-agent scenarios. Our research is supplemented by an analysis of the impact of the agent dynamics model on the intrinsic motivation module, insights into the performance of different CoHet variants, and its robustness to an increasing number of heterogeneous agents.
\end{abstract}


\newcommand{\BibTeX}{\rm B\kern-.05em{\sc i\kern-.025em b}\kern-.08em\TeX}

\begin{document}


\pagestyle{fancy}
\fancyhead{}


\maketitle 


\section{Introduction}
The paradigm of Multi-agent Reinforcement Learning (MARL) is rapidly emerging to be pivotal in a broad spectrum of practical applications such as resource management \citep{ying2005}, autonomous vehicles \citep{Cao2013Cars}, traffic signal control \citep{Calvo2018Traffic}, supply chain management \citep{fuji2018Supply}, robotics \citep{lillicrap2019multirobot}, robot swarms \citep{hüttenrauch2017RobotSwarms}, etc. These applications generally benefit from the efficient use of the diverse capabilities of heterogeneous agents. Moreover, the successful execution of tasks in these multi-agent systems requires the agents to adapt their behaviors to other agents for effective coordination rather than operating independently. The real-world deployment of these MARL systems typically involves the agents relying solely on the local environmental information and learning policies with infrequent environmental rewards \citep{Wiewiora2010,wakilpoor2020heterogeneous}.\par
Applications such as package transport \citep{Gerkey2002}, traffic lights control \citep{Calvo2018HeterogeneousMD}, disaster response \citep{Nathan2014}, agriculture \citep{ju2019}, etc. utilize agent heterogeneity such as distinct physical and behavioral traits of agents. Heterogeneity is also vital in multi-robot tasks as it enables efficient characterization and discovery of diverse behaviors, improving learning performance \citep{LiuLi2022}. On the other hand, the dependency on reward signals for the agent’s learning process introduces the issue of reward sparsity \citep{hare2019dealing}. Due to the lack of frequent feedback from the environment and the non-trivial nature of manually designing reward functions, MARL systems need to be robust enough to deal with infrequent environmental rewards. \par
\begin{figure*}[ht]
    \centering
    \includegraphics[width=.86\textwidth]{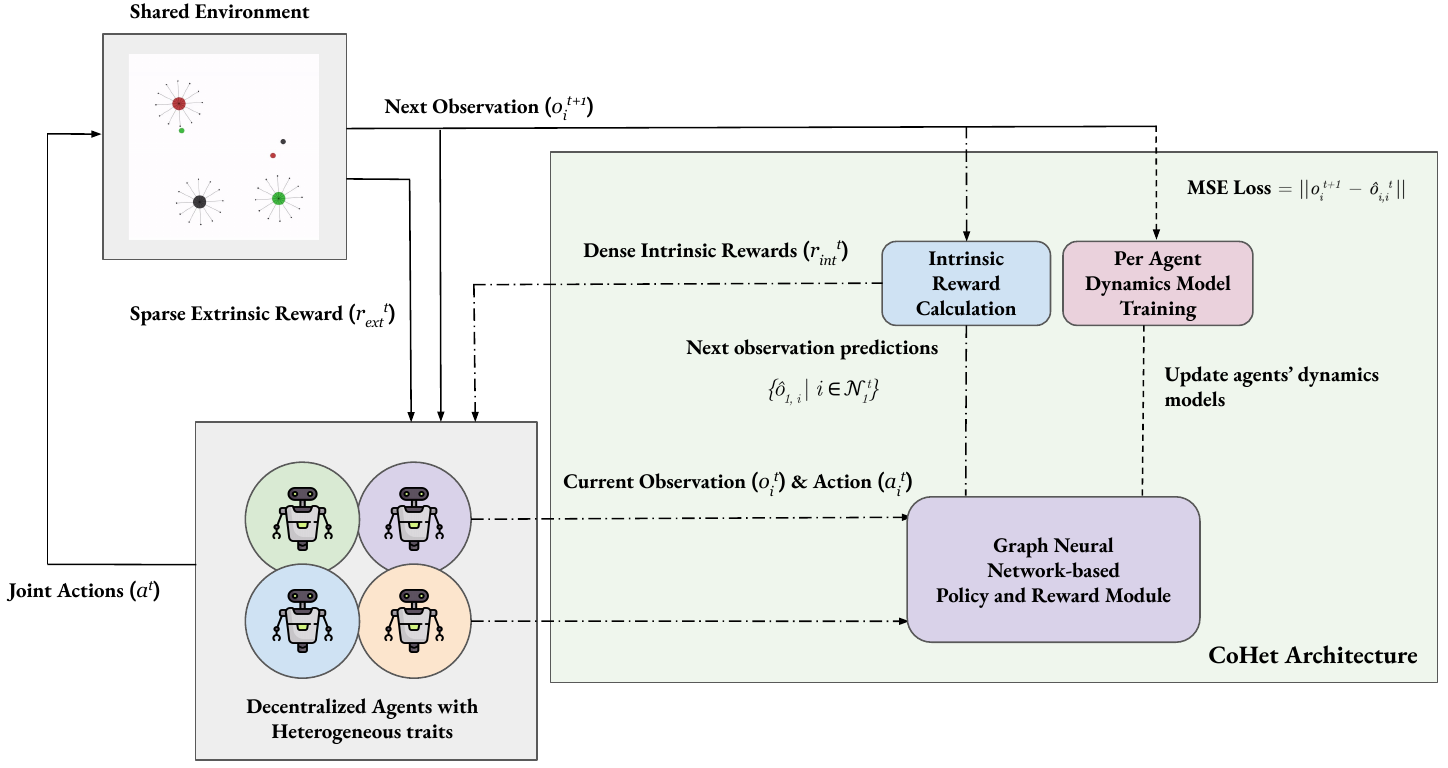}
    \caption{Overview of the CoHet intrinsic reward architecture: utilizing the observation predictions of neighboring agents, it augments the self-supervised intrinsic rewards with the sparse environmental rewards to elicit collaborative actions}
    \label{fig:cohet_overview}
    \Description{The figure gives an overview of the CoHet intrinsic reward calculation architecture – it shows how next observation predictions are utilized to calculate intrinsic rewards that are augmented with the extrinsic rewards.}
\end{figure*}
In addition to these challenges, the majority of real-world applications constrain the agents to act in a decentralized manner, and under partial observability, where each agent has a partial view of the shared environment. As a result, it is impractical for them to learn cooperative behaviors by utilizing a centralized algorithm that possesses global knowledge of all the agents and the state space \citep{iqbal19a,liu2019pic}. In comparison to centralized training and full observability, the challenges of agent heterogeneity and reward sparsity are more severe in decentralized training under partial observability \citep{ma2022elign}. Despite real-world requirements, existing solutions often rely on global parameter sharing or a centralized critic. \par
To the best of our knowledge, no prior research has addressed the issue of cooperative heterogeneous MARL in decentralized training settings under the practical constraints of real-world applications, such as partial observability and reward sparsity (see `Related Works' section for more details). We propose \textbf{CoHet}, an algorithm that facilitates heterogeneous agent cooperation addressing the constraints required for real-world deployments. CoHet does not require any prior knowledge of agent heterogeneity (e.g speed, size, type, agent index). It employs an architecture for learning heterogeneous MARL policies by utilizing a novel Graph Neural Network-based intrinsic motivation/reward calculation mechanism. In summary, our specific contributions are as follows:\par
\textbf{A Novel Intrinsic Reward Mechanism:} We introduce a novel self-supervised intrinsic reward calculation algorithm --- \textbf{CoHet}\footnote{The codebase for CoHet can be found at: \href{https://github.com/jahirsadik/CoHet-Implementation}{https://github.com/jahirsadik/CoHet-Implementation}}, utilizing the underlying communication graph of a Graph Neural Network (GNN). In comparison to previous methods, CoHet accurately estimates the intrinsic rewards in the presence of agent heterogeneity (e.g. physical attributes/composition, behavioral differences) by using only the agents' local neighborhood information. We present two formulations of the CoHet algorithm, one that utilizes the neighborhood predictions (CoHet\textsubscript{team}), and another that uses the agent's own predictions (CoHet\textsubscript{self}) for intrinsic reward calculation.\par
\textbf{Integration with Established Algorithms:} Our standalone intrinsic motivation architecture can be integrated with existing decentralized heterogeneous policy learning algorithms, thus enhancing performance in cooperative MARL benchmarks. We demonstrate this by incorporating the state-of-the-art HetGPPO algorithm \citep{bettini2023heterogeneous}, leveraging its underlying GNN communication graph for intrinsic reward calculation. In contrast to previous heterogeneous policy learning techniques, this formulation requires no prior knowledge of the types of agent heterogeneity, indexing, etc.

\textbf{Extensive Validation and Scalability:} We validate CoHet in the presence of heterogeneous agents in six different scenarios in the Multi-agent Particle Environment (MPE) and Vectorized Multi-Agent Simulator (VMAS) benchmarks, showing superior performance. We present findings on the impact of agent dynamics models on the intrinsic reward calculation, compare the two variants of the algorithm, and demonstrate its robustness to an increasing number of heterogeneous agents in a shared environment.\par
The subsequent sections will establish the necessary foundation of our work and explore our contributions in detail. In Section \ref{sec:related-works}, we present the related works in this domain and highlight the lack of studies addressing real-world constraints. In Section \ref{sec:bg-problem}, we formulate our problem and discuss the background necessary for our proposed method. Our proposed algorithm and system architecture are presented in Section \ref{sec:cohet-algorithm}. The experimental outcome of our research along with further studies has been delineated in Section \ref{sec:empirical_eval}. Finally, we conclude with an overview of our research and discuss the scope of potential future work in Section \ref{sec:conclusion}.
\section{Related Works}
\label{sec:related-works}
In response to the challenges posed by agent heterogeneity and the sparsity of rewards, most existing literature addresses one or the other. Existing methods addressing agent heterogeneity either necessitate prior knowledge of the types of heterogeneous agents \citep{foerster2016learning,Gupta2017CooperativeMC} or suggest solutions that are specific to problems within only certain sub-classes of heterogeneity \citep{wang2020roma,li2021celebrating,lowe2020multiagent}. Moreover, their use in partially observable systems is limited due to the lack of inter-agent communication. A notable study on heterogeneous MARL \cite{bettini2023heterogeneous}, proposes an algorithm termed HetGPPO (Heterogeneous GNN-based Proximal Policy Optimization), that is capable of learning heterogeneous decentralized policies in partially observable scenarios. It uses a GNN-based communication layer for sharing information among agents within local neighborhoods, thereby mitigating the effects of partial observability. Unlike previous methods, their reliance on only local information enables Decentralized Training with Decentralized Execution (DTDE). Similar to their work, CoHet utilizes only the local neighborhood information under the constraints of DTDE.\par
Although the aforementioned studies address the problem of agent heterogeneity, typically involving centralized critics, parameter sharing, or prior knowledge of agent heterogeneity, they do not consider inter-agent cooperation under reward sparsity. Dense scalar reward signals termed ``Intrinsic Motivation", are often used to encourage exploration or coordination among agents \citep{Yali2019,jaques19a,jeon2022maser}. 
In case of the algorithm proposed by \citet{ma2022elign} termed ELIGN (Expectation Alignment as a Multi-Agent Intrinsic Reward), the intrinsic rewards foster inter-agent coordination in a decentralized manner. However, a significant drawback of using an agent’s own dynamics model as a proxy to calculate neighborhood predictions in ELIGN is that it becomes more challenging for agents to accurately model the dynamics of other agents in the presence of agent heterogeneity. Inaccuracies in the dynamics model can result in misleading alignment signals, and as a result, ELIGN scales poorly in the presence of heterogeneous agents. In contrast, CoHet utilizes the local neighborhood information passed via the underlying GNN-communication graph, to more accurately model the heterogeneity among agents.\par
There are only a couple of existing research works that simultaneously tackle agent heterogeneity under constraints of partial observability and reward sparsity \citep{zheng2020cooperative,andres2022collaborative}. In the former work, heterogeneity is defined differently, referring to a mixture of on-policy, off-policy, and Evolutionary Algorithm (EA) agents and not the diverse physical and behavioral traits of agents. Moreover, their adoption of a local-global memory replay prevents them from undergoing training in a fully decentralized manner. The latter leverages intrinsic motivation to tackle heterogeneous agent cooperation under reward sparsity similar to our work. However, their utilization of a centralized critic that merges all agent parameters into a single network constrains its applicability in the DTDE paradigm. Despite its significance, the lack of solutions in decentralized training settings limits the deployment of MARL agents in practical applications. CoHet addresses this notable research gap in the area of cooperative heterogeneous MARL, by fostering cooperation in a decentralized manner, under the real-world challenges of partial observability and reward sparsity.
\section{Background}
\label{sec:bg-problem}
This section outlines the Markov games framework used to formulate our problem and the message-passing Graph Neural Network employed for differentiable inter-agent communication of local observations and predictions.
\subsection{Markov Games}
\label{subsec:markov-games}
The Markov games framework \citep{Littman_1994} is a generalization of a Markov decision process to the case of multiple agents with cooperating or competing goals. The Partially Observable Markov Games (POMG) framework is used under conditions of partial observability. It is defined using the tuple ---
$$\langle \mathcal{V}, \mathcal{S}, \mathcal{O}, \mathcal{A}, \{o_{i}\}_{i \in \mathcal{V}}, \{\mathcal{R}_{i}\}_{i \in \mathcal{V}}, \mathcal{T}, \gamma \rangle$$
where, \(\mathcal{V} = \{1, 2, \ldots, N\}\) represents the set of all \(N\) agents. The state space is denoted by \(\mathcal{S}\). The observation space is defined as \(\mathcal{O} \equiv \mathcal{O}_{1} \times \mathcal{O}_{2} \times \ldots \times \mathcal{O}_{n}\), where \(\mathcal{O}_{i} \subseteq \mathcal{S}\) for each agent \(i\) in \(\mathcal{V}\). The action space for all agents is denoted by \(\mathcal{A} \equiv \mathcal{A}_{1} \times \mathcal{A}_{2} \times \ldots \times \mathcal{A}_{n}\). Each agent \(i \in \mathcal{V}\) has an observation instance \(o_{i} \in \mathcal{O}_{i}\), which represents a partial view of \(\mathcal{S}\). The reward function for agent \(i\) is given by \(\mathcal{R}_{i} \colon \mathcal{S} \times \mathcal{A} \times \mathcal{S} \mapsto \mathcal{R}\). The stochastic state transition model is denoted by \(\mathcal{T} \colon \mathcal{S} \times \mathcal{A} \times \mathcal{S} \mapsto [0, 1]\), describing how each agent transitions to the next state, given the current state and the action taken. Finally, \(\gamma\) represents the discount factor, where \(0 \leq \gamma \leq 1\). Additionally, the reward that agent $i$ receives at time step $t$ can be denoted using $r_{i}^{t}=\mathcal{R}_{i}(s^{t},a^{t},s^{t+1})$, where $a^{t}=(a_{1}^{t},a_{2}^{t},\ldots,a_{n}^{t})$ is the joint action taken by all the agents at time step $t$. The goal of each agent is to maximize the total expected discounted return $R_{i}=\sum^{T}_{t=0}\gamma^{t}r_{i}^{t}$ over the course of an episode with horizon $T$. 
\subsection{Message-passing Graph Neural Network}
\label{subsec:message-passing-gnn}
The message-passing technique \citep{kipf2017semisupervised} used by GNNs to transfer information from one node to another has proven to be an effective learning framework for understanding the patterns, the neighborhood of nodes, and the sub-graphs in large graphs \citep{zhou2021graph}. In our problem formulation, we define graph $\mathcal{G}=(\mathcal{V},\mathcal{E})$, where
\begin{itemize}
    \item $\mathcal{V}$ is the set of vertices we use to represent the agents
    \item $\mathcal{E}$ is the set of edges discovered by inserting an edge $(i,j)$ from $i$ to $j$, if agent $j$ is within the observation radius of agent $i$
    \item $x_i$ represents the agent (i.e., node) attributes for each agent $i \in \mathcal{V}$. In our setting, this includes the non-absolute features of agent observations, which are found by removing the absolute features such as agent position, velocity, etc., from the agent observation $o_i$.
    \item $e_{ij}$ are the edge attributes for each edge $(i,j) \in \mathcal{E}$. The absolute position and the velocity of each agent are used to calculate the relative position and velocity. These are then concatenated to be used as the edge features.
\end{itemize}
For the message-passing step in our GNN, at each time step, the agent embeddings and edge features are first computed and then utilized in the message-passing GNN kernel to learn the local sub-graph. Information is iteratively passed between adjacent agents along the edges of the graph structure. The use of only the non-absolute observation features as agent embeddings allows the outputs of the message-passing GNN kernel to be invariant to geometric translations, thereby enhancing generalization. In order to incorporate the CoHet architecture on top of the underlying GNN formulation of HetGPPO, the inputs to our message-passing GNN kernel consist of the agent embedding $z_i = \omega_{\theta_i}(x_i)$ and edge attributes $e_{ij} = p_{ij} \| v_{ij}$. Here, $\omega_{\theta_i}$ represents a Multi-Layer Perceptron (MLP) encoder with parameters $\theta_i$, and $x_i$ represents the non-absolute features of the agent obtained by removing absolute geometric features from the observation $o_i$. Edge attributes $e_{ij}$ are calculated using the absolute features of agent observations, such as position $p_{ij}$ and relative velocity $v_{ij}$. Using local information from all neighbors $j \in \mathcal{N}_i$, the GNN model output $h_i$ for agent $i$ is calculated in Equation \ref{eqn:hetgppo}.

\begin{equation}
    \label{eqn:hetgppo}
    h_i = \psi_{\theta_i}(z_i) + \bigoplus_{j \in \mathcal{N}_i} \phi_{\theta_i}(z_j \| e_{ij})
\end{equation}
In Equation \ref{eqn:hetgppo}, $\psi_{\theta_i}$ and $\phi_{\theta_i}$ are two MLPs parameterized by $\theta_i$ and the aggregation operator $\bigoplus$ sums the $\phi_{\theta_i}$ outputs for all the neighbors of agent $i$. Finally, two distinct MLP decoders take the GNN output $h_i$ and produce the value $V_i(o_{\mathcal{N}_i})$ and the action $a_i$, distributed according to $a_i \sim \pi_i(\cdot\mid o_{\mathcal{N}_i})$. This formulation of GNN allows us to utilize it for both intrinsic reward calculation and heterogeneous policy learning of HetGPPO, based on local neighborhood information.


\begin{algorithm}[b]
\caption{CoHet Algorithm}
\label{alg:cohet}
\begin{algorithmic}[1]
\State Initialize models $\omega_{i}, \psi_{i}, \phi_{i}, \Omega_{i}, \Gamma_{i}, f_{i}$ with random values $\theta_i$, where $i \in \{1, 2, \ldots, N\}$\label{line:initialization}
\For{$k = 1, 2, \ldots$} \label{line:iter-start}
    \State Initialize set of trajectories for all agents, $\mathcal{D}_k \gets \{\}$ 
    \Statex \textsc{// Action \& Value calculation} 
    \For{$t = 0, 1, \ldots, T$}
        \For{$i = 1, 2, \ldots, N$} 
            \State $x_i^t \gets trim(o_{i}^t)_{\{p_{i}^t, v_{i}^t\}}$ \label{line:non-abs-feature}
            \State $z_i^t \gets \omega_{\theta_{i}}(x_i^t)$ \label{line:non-abs-emb}
            \State $h_i^t \gets \psi_{\theta_{i}}(z_i^t)$
            \For{each $j \in \mathcal{N}_{i}^{t}$} \label{line:h-start}
                \State $e_{ij}^t \gets p_{ij}^t \| v_{ij}^t$ \label{line:abs-emb}
                \State $h_{i}^t \gets h_{i}^t + \bigoplus \phi_{\theta_{i}}(z_j^t \| e_{ij}^t)$ \label{line:gnn-output}
            \EndFor \label{line:h-end}
            \State $a_i^t \gets \Omega_{\theta_{i}}(h_i^t)$ \label{line:out-act}
            \State $V_i^t \gets \Gamma_{\theta_{i}}(h_i^t)$ \label{line:out-val}
        \EndFor
        \State $a^t \gets a^t_1 \| a^t_2 \| \ldots \| a^t_N$ \label{line:concat-action}
    \EndFor
        \Statex \textsc{// Intrinsic reward calculation} 
    \For{$t = 0, 1, \ldots, T$}
        \For{$i \in \{1, 2, \ldots, N\}$} 
            \State $r_{int_{i}}^t \gets 0$ \label{line:intrew-start}
            \For{each $j \in \mathcal{N}_{i}^{t} \cap \mathcal{N}_{i}^{t+1}$} \label{line:rint-start}
                \State $w_j \gets \frac{d(i,j)}{\sum_{k \in \mathcal{N}_i^t \cap \mathcal{N}_i^{t+1}}{d(i,k)}}$ \label{line:rew-weight}
                \State $r_{int_{i}}^t \gets r_{int_{i}} + w_{j} \times -\|o^t_{t+1} - f_{\theta_{j}}(o_{i}^{t}, a_{i}^{t})\|$ \label{line:intrew-calc}
            \EndFor
            \State $r_{total_{i}}^t \gets r_{ext_{i}}^t + \beta \times r_{int_{i}}^t$ \label{line:intrew-end}
        \EndFor
        \State $r^t \gets r^t_{total_1} \| r^t_{total_2} \| \ldots \| r^t_{total_N}$ \label{line:concat-reward}
        \State $\mathcal{D}_k \gets \mathcal{D}_k \cup (o^t, a^t, r^t, o^{t+1})$ \label{line:collect-trajectory}
    \EndFor
    \State Use $\mathcal{D}_k$ to for Multi-PPO policy optimization\label{line:iter-end}
\EndFor
\end{algorithmic}
\end{algorithm}
\section{The CoHet Algorithm}
\label{sec:cohet-algorithm}
Real-world deployment of multi-agent systems requires agents that can deal with challenges such as decentralized training, operating with a partial view of the environment, and learning from infrequent environmental feedback signals. In this section, we introduce CoHet, a decentralized algorithm designed to enhance cooperation among heterogeneous agents in partially observable environments with sparse rewards. It provides a standalone self-supervised intrinsic reward architecture that can be incorporated with existing decentralized policy optimization algorithms. It fosters the learning of collaborative behaviors by reducing future uncertainty within each agent’s neighborhood. CoHet encourages the agents to align their actions with their neighbors’ predictions by imposing intrinsic reward penalties that deter deviations from such alignment. Furthermore, these calculated rewards serve as a source of dense reward signals that facilitate policy learning in numerous real-world tasks where manually designing reward functions is infeasible. 

\subsection{Algorithm Description}
\label{subsec:cohet-arch}
\begin{figure}[ht]
    \centering
    \begin{subfigure}[b]{0.4\textwidth}
        \centering
        \includegraphics[width=\textwidth]{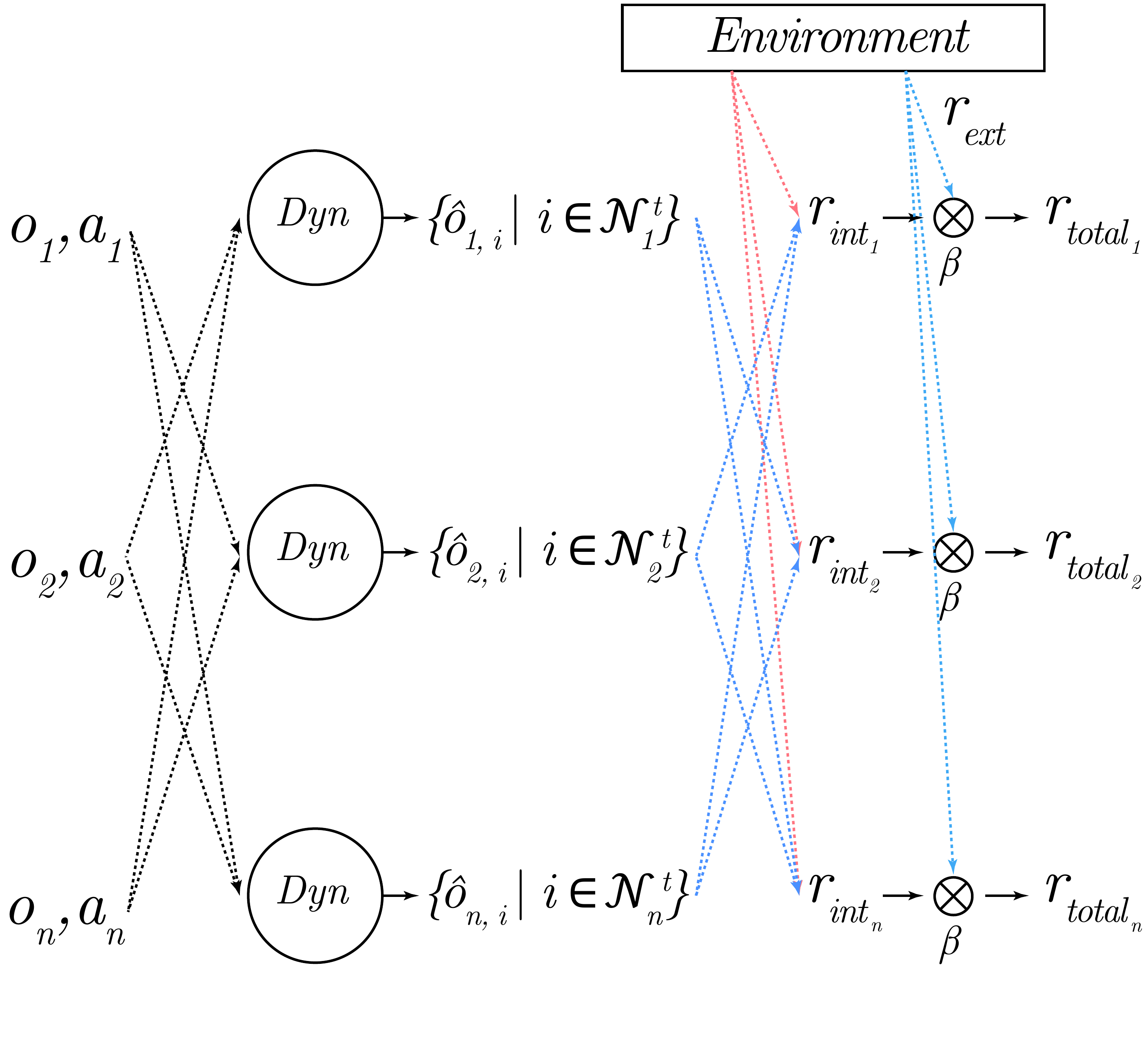}
        \caption{Reward calculation architecture}
        \label{fig:dyn-model-diagram}
    \end{subfigure}
    \hfill
    \begin{subfigure}[b]{0.4\textwidth}
        \centering
        \includegraphics[width=\textwidth]{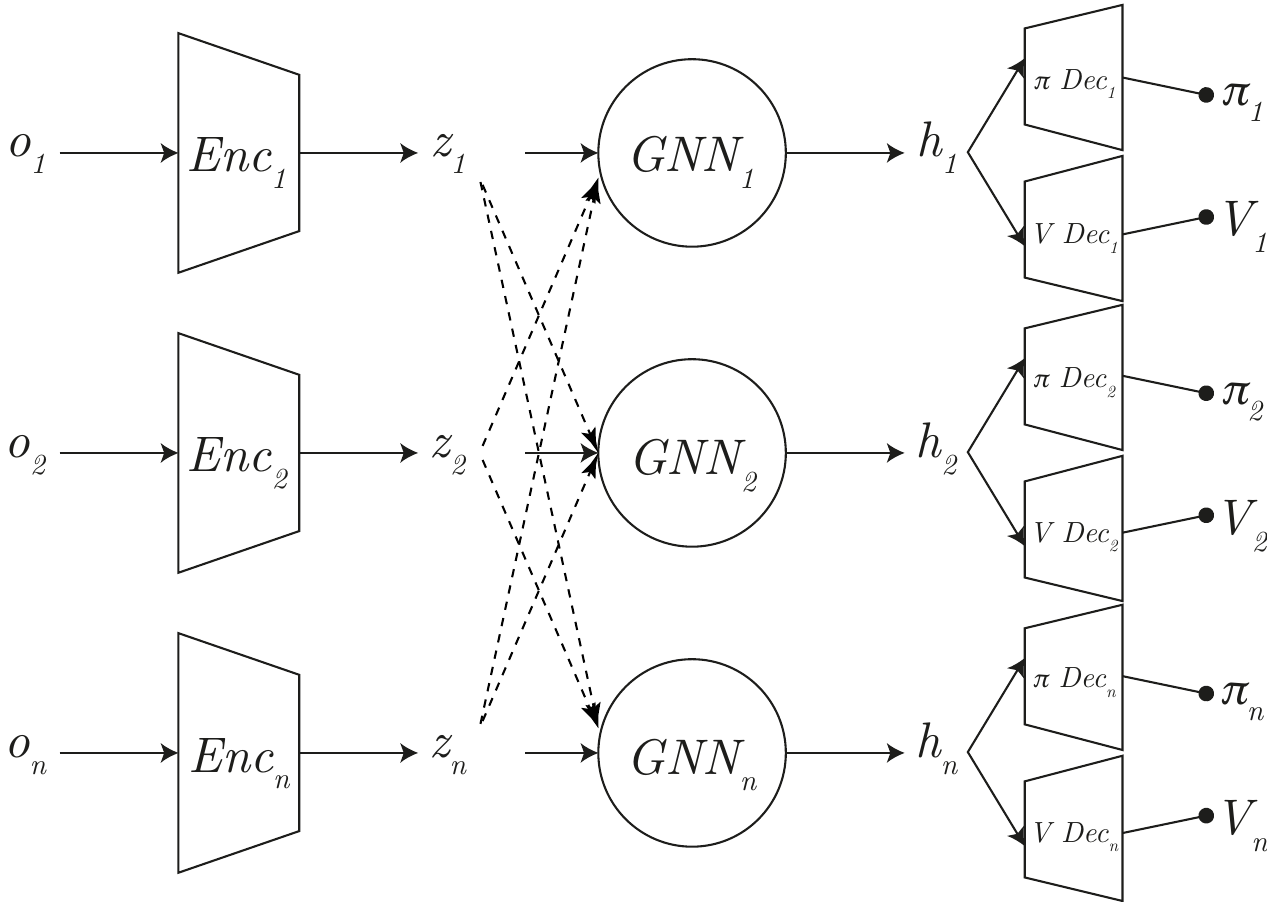}
        \caption{HetGPPO policy learning}
        \label{fig:hetgppo-diagram}
    \end{subfigure}
    \caption{The per-agent dynamics models in Figure \ref{fig:dyn-model-diagram} are used for calculating the intrinsic rewards, which are then combined with the extrinsic reward from the environment, resulting in $r_{total_i}$ for each agent $i$. This combined reward is passed to the HetGPPO policy learning module in Figure \ref{fig:hetgppo-diagram} for heterogeneous policy learning}
    \label{fig:combined}
    \Description{Figure showing two subfigures. Subfigure (a) depicts the reward calculation architecture, illustrating how intrinsic rewards are computed for each agent. Subfigure (b) shows the HetGPPO policy learning process, demonstrating how the combined reward, which includes both intrinsic and extrinsic components, is used to train the policy for heterogeneous agents.}
\end{figure}

The CoHet algorithm utilizes an underlying communication graph $\mathcal{G} = (\mathcal{V}, \mathcal{E})$ for passing both the ground truth observation $o_i^t$ of agent $i$ at time $t$, and its predicted next observation set $\{\hat{o}_{i,j}^t \mid j \in \mathcal{N}_i^{t+1}\}$ to all its local neighbors $\mathcal{N}_i^{t+1}$ at the next time step. As previously mentioned, CoHet's standalone reward calculation architecture presented in Figure \ref{fig:dyn-model-diagram} can be used alongside existing decentralized multi-agent policy optimization algorithms. However, integrating CoHet’s decentralized heterogeneous intrinsic motivation architecture with established policy learning frameworks like HetGPPO, which accommodates policy heterogeneity in MARL scenarios, can be advantageous for deploying agents with varied physical or behavioral characteristics, such as varying sizes, speeds, action spaces, etc. Hence, in Figure \ref{fig:hetgppo-diagram}, we demonstrate the incorporation of the HetGPPO policy optimization architecture.\par
In Algorithm \ref{alg:cohet}, we start by initializing the model for each agent which includes the encoder $\omega_{i}$, two multi-layer perceptrons (MLPs) $\psi_{i}$ and $\phi_{i}$, $\pi$-decoder $\Omega_{i}$, value decoder $\Gamma_{i}$, and dynamics model $f_{i}$. At each training iteration, the observations $o_i^t$ are collected and the position $p_i^t$ and velocity $v_i^t$ are trimmed from it to obtain the non-absolute features $x_i^t$ in Line \ref{line:non-abs-feature}. These are used to calculate the node embedding $z_i^t$ in Line \ref{line:non-abs-emb}. The position $p_i^t$ and velocity $v_i^t$ from the observations are used as edge embeddings in Line \ref{line:abs-emb}, which allows for GNN outputs to be invariant to geometric translations. Neighboring agent embeddings $\{z_j^t \mid j \in \mathcal{N}_i^t\}$ are obtained through the GNNs underlying differentiable communication channel. These neighborhood embeddings, along with edge features are aggregated to calculate the GNN model output in Line \ref{line:gnn-output}, which is then decoded to produce action and value outputs in Line \ref{line:out-act} and Line \ref{line:out-val}. The joint action $a^t$, calculated in Line \ref{line:concat-action}, and the individual agent actions $a^t_i$ are later used for policy optimization and intrinsic reward calculation.
The procedure for policy optimization as discussed above is visualized in Figure \ref{fig:hetgppo-diagram}, where the reward assigned to the agents at each time step is augmented with the dense intrinsic reward calculated via the process outlined in Figure \ref{fig:dyn-model-diagram}. This augmentation of dense intrinsic reward signals allows the agents to learn from the interactions that do not result in any extrinsic/environmental rewards.\par
\begin{figure}[h!]
    \centering
    \includegraphics[width=0.46\textwidth]{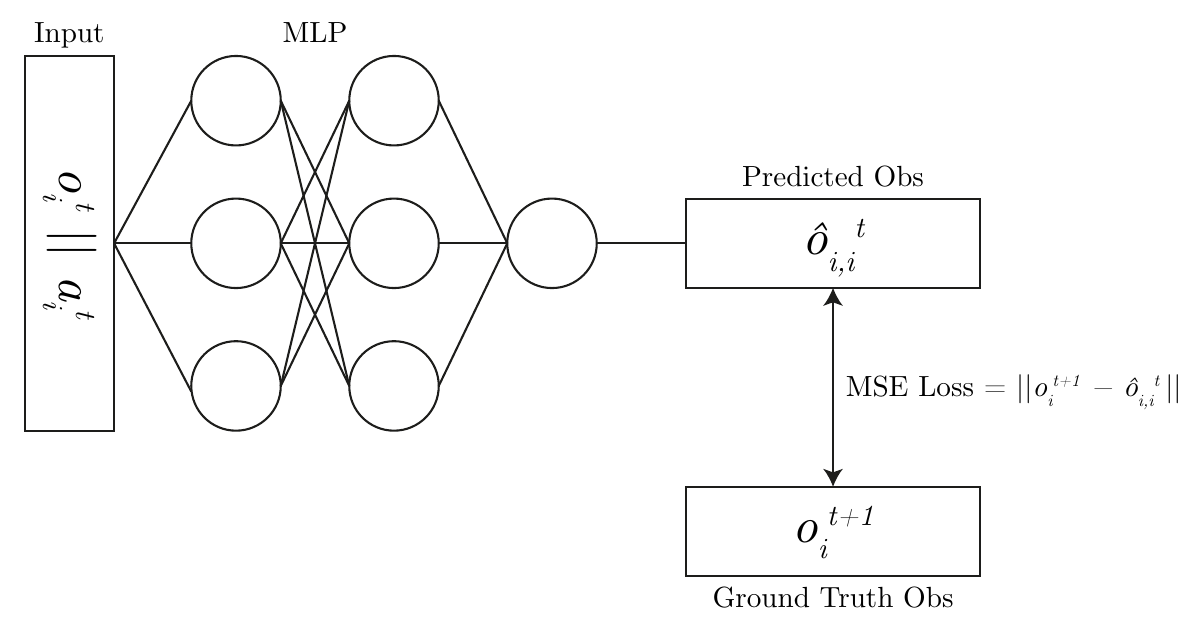}
    \caption{Agent dynamics model training process utilizes the ground truth next observation at the next timestep to train the agents to predict next observations more accurately}
    \label{fig:training}
    \Description{Figure depicting the agent dynamics model training process. The figure shows the workflow or steps involved in training the dynamics model for agents, including any key stages or components of the process. It illustrates the various phases of training and interactions between different elements of the model. This helps in understanding how the dynamics model is developed and refined.}
\end{figure}
In the CoHet algorithm, the dynamics model $f_{\theta_i}(o_i^t, a_i^t)$ of each agent $i$, are continually trained (Line \ref{line:intrew-start} to Line \ref{line:intrew-calc}) using the agents own experiences in the environment. This is done in order to be used for intrinsic reward calculation. Agents achieve familiarity with environmental dynamics through continuous training in the dynamics model, which they use to predict the next observation for their neighbors. As visualized in Figure \ref{fig:training}, during the training of the dynamics model, the observation $o_i^t$, and the action $a_i^t$ are passed as input and the model outputs the prediction of the next observation $\hat{o}_{i,i}^t$, where $\hat{o}_{i,i}^t$ refers to the predicted next observation by agent $i$ for itself. We employ a three-layer MLP with ReLU non-linearities as the dynamics model and train them to minimize the mean squared error (MSE) between its prediction and the ground truth next observation $o_i^{t+1}$.\par

In addition, instead of giving equal weight to each neighbor’s prediction in the calculation of the intrinsic reward, we utilize Euclidean distance-based weighting to prioritize the predictions of the agents in close proximity. To achieve this, we calculate the inverse of the Euclidian distance between agent $i$ and agent $j$, obtained from their positions $p_i$ and $p_j$ in Equation \ref{eqn:cohet:dist} and use it to calculate the weights for each neighbor in Equation \ref{eqn:cohet:weight}. Note that, the agent dynamics model predictions for the neighbors are communicated in the next time step, so we only consider the agents present in the communication neighborhood at both time $t$, and $t+1$, denoted by $\mathcal{N}_{i}^{t} \cap \mathcal{N}_{i}^{t+1}$.\par

\begin{equation}
    \label{eqn:cohet:dist}
    d(i, j)=(\|p_i - p_j\|)^{-1}
\end{equation}

\begin{equation}
    \label{eqn:cohet:weight}
    w_j=\frac{d(i,j)}{\sum_{k\in \mathcal{N}_i^t \cap \mathcal{N}_i^{t+1}}{d(i,k)}}
\end{equation}

\begin{equation}
    \label{eqn:cohet:pred}
    \hat{o}_{j,i}^t = f_{\theta_j}(o_i^t, a_i^t)
\end{equation}

\begin{equation}
    \label{eqn:cohet:rew}
    r_{int_i}^t(o_i^t, a_i^t) = - \sum_{j\in \mathcal{N}_i^t \cap \mathcal{N}_i^{t+1}}{w_j \times \|o_i^{t+1} - \hat{o}_{j,i}^t\|}
\end{equation}

The predicted next observation for agent $i$ by each of its neighbor $j$ that are present at both time $t$ and $t+1$ ($j \in \mathcal{N}_{i}^{t} \cap \mathcal{N}_{i}^{t+1}$), is calculated using their dynamics model $f_{\theta_j}$ in Equation \ref{eqn:cohet:pred}. These predictions are subsequently used to calculate the weighted intrinsic rewards in Equation \ref{eqn:cohet:rew}. Here, the misalignment term $\|o_i^{t+1} - \hat{o}_{j,i}^t\|$ is the absolute error between the ground truth observation of agent $i$ at time $t+1$ and the predicted next observation by its neighbors at time $t$. It essentially represents the misalignment of agent $i$ with agent $j$'s predictions. The additive inverse is thus taken to impose penalties or negative rewards for misalignment. Finally, the intrinsic reward is multiplied by a hyperparameter $\beta$ and added to the extrinsic rewards. This procedure is visualised on the left side of Figure \ref{fig:dyn-model-diagram}. Finally, the obtained total rewards $r_{total_{i}}^t$ for each agent, as depicted in the right side of Figure \ref{fig:dyn-model-diagram}, are the dense reward signal which can now be used for policy optimization. Per-agent total rewards $r_{total_{i}}^t$, and the concatenated rewards $r^t$ are calculated, and the collected trajectories are then used for policy optimization from Lines \ref{line:intrew-end} to \ref{line:iter-end}.\par 
This method of reward computation using only the local neighborhood predictions, as discussed above, is termed \textbf{CoHet\textsubscript{team}}. Here, each agent effectively computes a prediction of the next observation for each of its neighboring agents at time $t$ and communicates that prediction at the next time step. This procedure results in each agent gathering the predictions of its neighbors. To coordinate their actions based solely on their local environment, the agents need to adapt their behaviors and actions, to the predictions of the next observations made by their neighbors. In order to foster the learning of this coordination, the intrinsic reward signal is calculated as a penalty for misalignment by taking the negative of the absolute prediction error. Agent policies are subsequently optimized to maximize the overall rewards and thus minimize these intrinsic reward penalties, encouraging them to adapt to neighborhood predictions over time. We also propose another variant (\textbf{CoHet\textsubscript{self}}) of the algorithm, which simply passes the ground truth observation and the action of the agent to its own dynamics model for prediction, resulting in each agent only following its own predictions independently. \par
To summarize, at each time step, the architecture of CoHet calculates dense intrinsic reward signals and augments those to the sparse environmental rewards to optimize policies for heterogeneous agents while relying solely on local communication and local observations. This allows the training of agents in a decentralized manner, consistent with the DTDE paradigm. 

\section{Experiments}
\label{sec:empirical_eval}
In this section, we demonstrate that both variants of CoHet (CoHet\textsubscript{self}, CoHet\textsubscript{team}) outperform the state-of-the-art decentralized heterogeneous MARL policy learning algorithm HetGPPO in each of the tasks evaluated on widely used VMAS and MPE benchmarks. The incorporation of the CoHet architecture leads to the learning of collaborative behaviors among heterogeneous agents, as evidenced by the improved performance over the baseline in these cooperative MARL tasks. We additionally compare with the state-of-the-art MARL baseline, IPPO (Independent Proximal Policy Optimization), which is applicable in decentralized training settings for heterogeneous agents under partial observability similar to HetGPPO. Unlike the two centralized critic-based heterogeneous MARL approaches discussed in the `Related Works' section or widely used algorithms such as MADDPG \cite{lowe2020multiagent}, MAPPO \cite{yu2022surprisingeffectivenessppocooperative}, COMA \cite{foerster2017counterfactualmultiagentpolicygradients}, etc., these baselines along with CoHet address the more challenging problem of not relying on any centralized controller or prior knowledge of agent heterogeneity, but rather optimizing policies based on only the locally observable partial information available to these heterogeneous agents. As a result, to maintain uniform assumptions across methods, we show comparisons with the existing decentralized heterogeneous algorithms that operate under similar constraints. Each of our evaluated tasks involves agents trained in a fully decentralized manner following the principles of the DTDE paradigm, acting under partial observability and reward sparsity. Furthermore, we analyze how each agent learns the dynamics model as time progresses and how it results in the reduction of the intrinsic reward penalty for misalignment. We compare the two variants of CoHet (CoHet\textsubscript{team} and CoHet\textsubscript{self}) and how they perform. We also demonstrate that the CoHet algorithm is robust to an increasing number of heterogeneous agents in the shared environment, an issue previously encountered in the intrinsic motivation-based methods \cite{ma2022elign}.
\subsection{Environments}
\label{subsec:environments}
For our experiments, we choose the MPE and VMAS settings that are extensively used for the performance evaluation of MARL agents in a variety of cooperative tasks \citep{lowe2020multiagent,bettini2023snd,bou2023torchrl}. We evaluate across six diverse scenarios, where heterogeneous agents have to cooperate to achieve a shared goal under partial observability ---
\begin{itemize}
    \item \textbf{VMAS Flocking:} $N$ agents flock around a landmark with random obstacles. Higher velocity and smaller inter-agent distances are rewarded. Collisions are penalized.
    \item \textbf{MPE Simple Spread:} $N$ agents aim to occupy $N$ distinct landmarks, receiving rewards for doing so and incurring penalties for collisions.
    \item \textbf{VMAS Reverse Transport:} $N$ agents collaboratively push a heavy package toward a random goal within an enclosure, with rewards given upon reaching the goal.
    \item \textbf{VMAS Joint Passage:} Two agents, connected by an asymmetric linkage, navigate through a narrow passage to reach a goal on the other side of a wall.
    \item \textbf{VMAS Navigation:} Multiple landmarks are color-mapped to heterogeneous agents. Each agent must reach its designated color landmark to maximize rewards.
    \item \textbf{VMAS Sampling:} $N$ agents start in a grid environment with one-time rewards. Heterogeneous agents use LIDARs with varying ranges and face partial observability.
\end{itemize}

\begin{figure*}[!t]
\captionsetup[subfigure]{justification=Centering}
\begin{subfigure}[t]{.33\textwidth}
    \includegraphics[width=\textwidth,height=0.2\textheight]{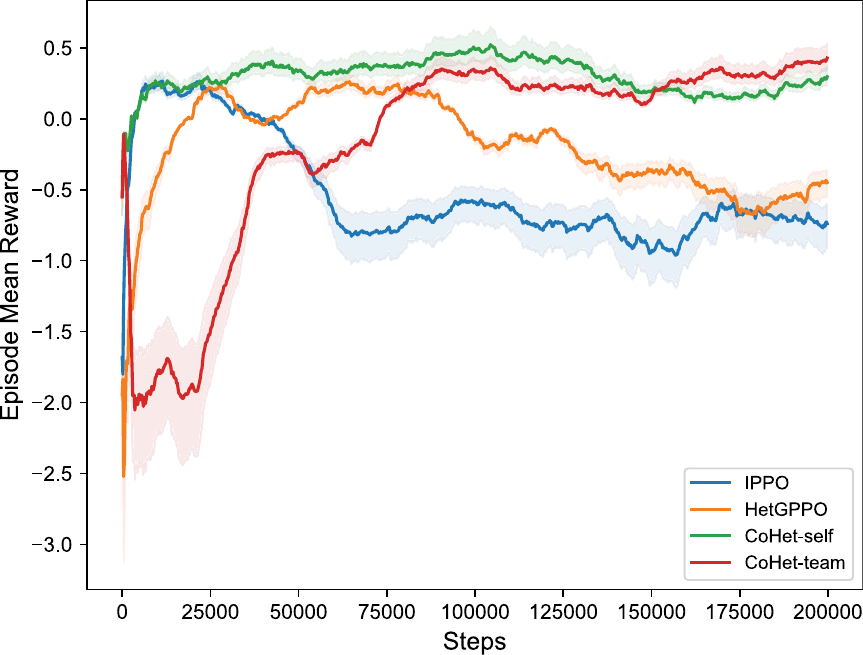}
    \caption{VMAS Flocking}
    \label{subfig:flocking1}
\end{subfigure}
\begin{subfigure}[t]{0.33\textwidth}
    \includegraphics[width=\textwidth,height=0.2\textheight]{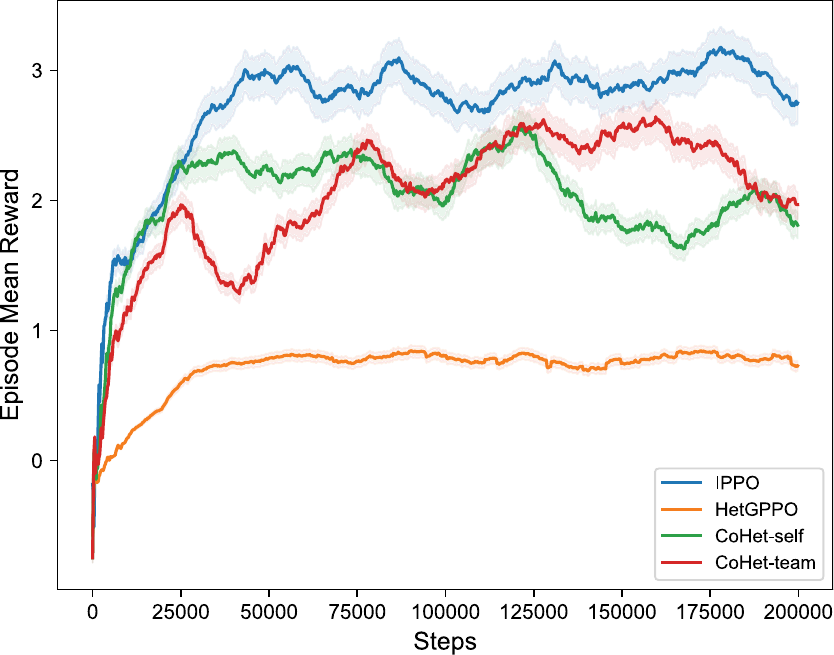}
    \caption{VMAS Navigation}
    \label{subfig:navigation1}
\end{subfigure}
\begin{subfigure}[t]{0.33\textwidth}
    \includegraphics[width=\textwidth,height=0.2\textheight]{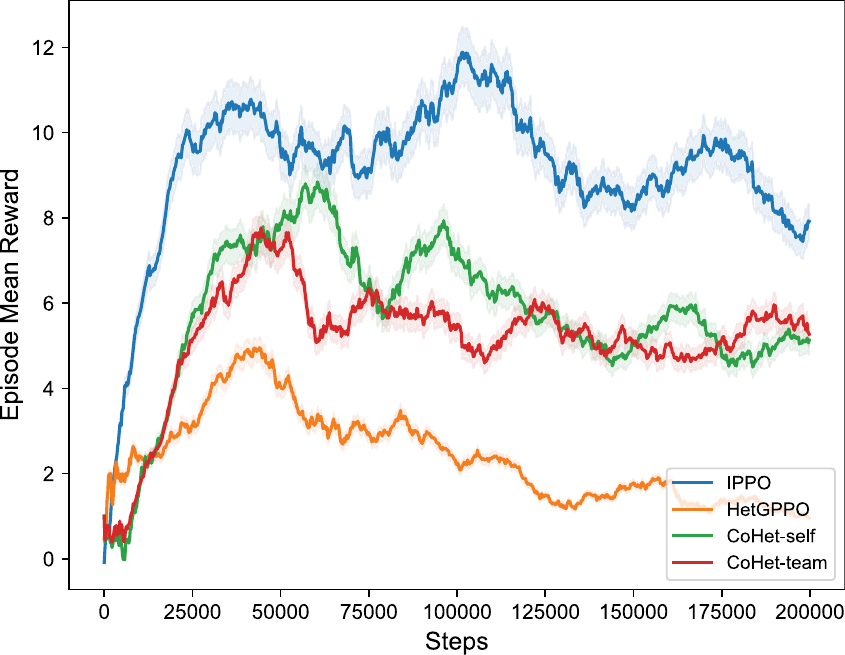}
    \caption{VMAS Reverse Transport}
    \label{subfig:rev-transport1}
\end{subfigure}
\begin{subfigure}[t]{.33\textwidth}
    \includegraphics[width=\textwidth,height=0.2\textheight]{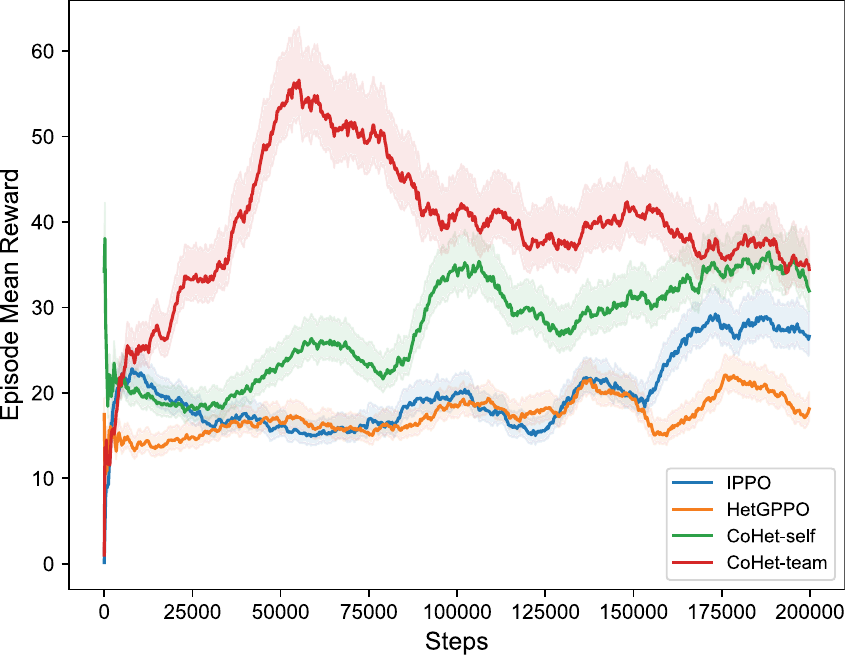}
    \caption{VMAS Sampling}
    \label{subfig:sampling1}
\end{subfigure} \hfill
\begin{subfigure}[t]{0.33\textwidth}
    \includegraphics[width=\textwidth,height=0.2\textheight]{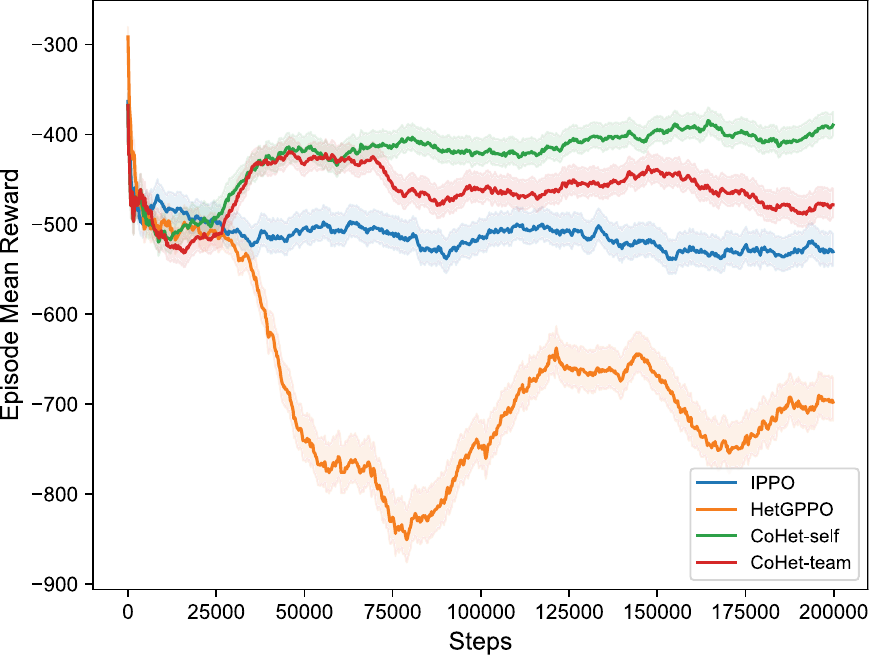}
    \caption{MPE Simple spread}
    \label{subfig:simple-spread1}
\end{subfigure} \hfill
\begin{subfigure}[t]{0.33\textwidth}
    \includegraphics[width=\textwidth,height=0.2\textheight]{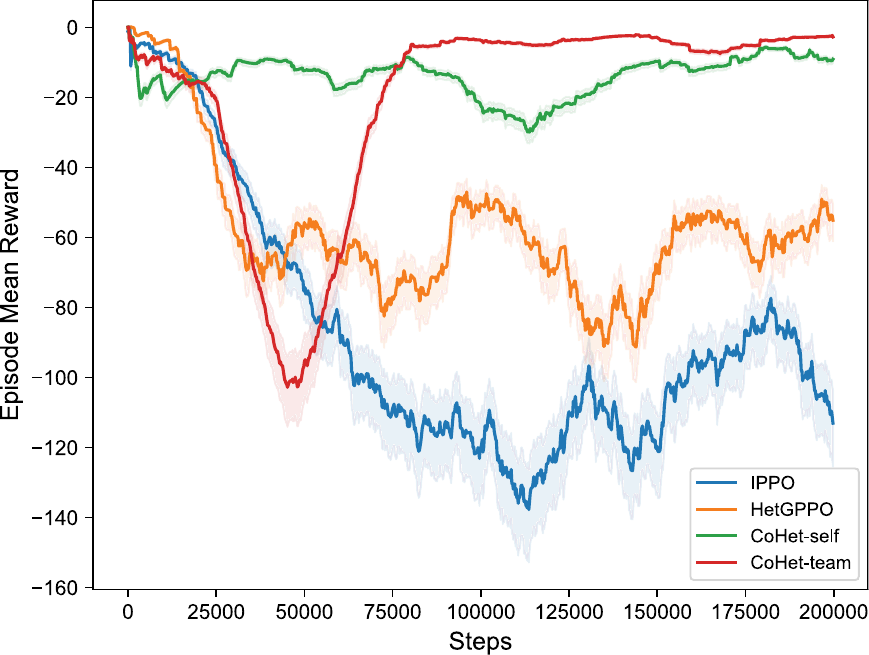}
    \caption{VMAS Joint Passage}
    \label{subfig:joint-passage-hetgppo1}
\end{subfigure}
\caption{Mean Episodic Rewards in VMAS and MPE cooperative multi-agent benchmarks demonstrate that in each of the scenarios, both variants of CoHet (self/team) outperform the HetGPPO baseline, and outperform IPPO in four out of six tasks} 
\label{fig:mean-extrinsic-rewards}
\Description{Figure showing the Mean Episodic Rewards in various cooperative multi-agent benchmarks. The figure is divided into six subfigures, each representing a different task or environment: (a) VMAS Flocking, (b) VMAS Navigation, (c) VMAS Reverse Transport, (d) VMAS Sampling, (e) MPE Simple Spread, and (f) VMAS Joint Passage. Each subfigure displays the performance of both CoHet variants (self and team) compared to the HetGPPO baseline. The results indicate that CoHet variants consistently outperform the HetGPPO baseline across all tasks, and they outperform IPPO in four out of six tasks, demonstrating a significant improvement in agent performance in these multi-agent scenarios. The figure provides a comprehensive overview of the effectiveness of CoHet in diverse environments.}
\end{figure*}
\subsection{Empirical Results}
\label{subsec:results}
CoHet outperforms HetGPPO in each of the six cooperative scenarios involving agents with heterogeneous traits (e.g. size, speed, observation radius). Moreover, both the CoHet\textsubscript{team} and CoHet\textsubscript{self} variants outperform the baseline in each of these cases. These results indicate that incorporating the GNN-based intrinsic motivation of CoHet enables the agents to surpass the baseline performance. In case of the two different variants of the algorithm, CoHet\textsubscript{self} only outperforms the rest in the MPE Simple Spread scenario, as shown in Figure \ref{subfig:simple-spread1}. On the other hand, CoHet\textsubscript{team} outperforms in the rest of the scenarios -- VMAS Navigation, VMAS Reverse Transport, VMAS Sampling, VMAS Flocking, and VMAS Joint Passage, as depicted by the Figures \ref{subfig:flocking1}, \ref{subfig:navigation1}, \ref{subfig:rev-transport1}, \ref{subfig:sampling1}, and \ref{subfig:joint-passage-hetgppo1}, respectively. Differing convergence properties of the agent dynamics models required a different number of training steps in each of these environments. All the environments in Figure \ref{fig:mean-extrinsic-rewards} were trained with a train batch size of 60000, in 4 random environment initializations, and for 1000 episodes where each episode lasted for a maximum of 200 environmental steps on a machine with an NVIDIA GeForce RTX 3090 GPU. For reproducibility of all our research, we include the details about the model hyper-parameters, and environment parameters in the supplementary materials. The code will be made public upon acceptance. \par
We hypothesize that the enhanced performance of CoHet over HetGPPO in the multi-agent cooperative tasks of Figure \ref{fig:mean-extrinsic-rewards} arises due to the effect of the intrinsic rewards acting as a penalty for misalignment with neighborhood predictions, thereby encouraging the agents to behave in a manner that reduces future uncertainty for their neighborhood. Moreover, the dense intrinsic rewards that foster collaborative actions in local neighborhoods also guide the agents during interactions with infrequent/sparse environmental feedback. Whereas CoHet allows agents to perform coordinated exploration even under reward sparsity, HetGPPO suffers when environmental (i.e. extrinsic) rewards are infrequent.\par
\begin{table}[h!]
\centering
\caption{Mean Episodic Reward of CoHet variants vs. state-of-the-art baselines after \( 2 \times 10^5 \) environment steps. Both CoHet variants simultaneously outperform the HetGPPO baseline in each task and outperform Heterogeneous Independent PPO (IPPO) in four out of six tasks}
\label{tab:comparisons-to-ippo}
\begin{tabular}{@{}lcccc@{}} \toprule
\textbf{Scenario} & \textbf{IPPO} & \textbf{HetGPPO} & \textbf{CoHet\textsubscript{team}} & \textbf{CoHet\textsubscript{self}} \\ \midrule
Flocking & -0.73 & -0.49 & \textbf{0.41} & 0.28 \\ 
Navigation & \textbf{2.93} & 0.75 & 1.97 & 1.80 \\ 
Rev. Trans. & \textbf{7.92} & 0.96 & 5.27 & 5.13 \\ 
Sampling & 26.13 & 17.81 & \textbf{34.86} & 31.75 \\ 
Sim. Spread & -528.98 & -701.15 & -477.73 & \textbf{-390.18} \\ 
Joint Pass. & -112.47 & -55.10 & \textbf{-2.73} & -9.11 \\ \bottomrule
\end{tabular}
\end{table}
We further evaluate our method against another state-of-the-art, Independent Proximal Policy Optimization (IPPO), where each agent acts independently in the shared environment and optimizes their policies using PPO. IPPO has been demonstrated to outperform many fully observable critic models on several MARL benchmarks \citep{dewitt2020independent}. However, a drawback of using IPPO is that the other agents are considered as environmental components and are not explicitly represented in the IPPO critic. Conditioning the critic on local agent observations rather than the entire state also leads to non-stationarity during the training \citep{bettini2023heterogeneous}. Despite potential difficulties, it has the positive aspect of not requiring prior knowledge of agent heterogeneity or global information during training, allowing it to be a suitable baseline for our evaluation. In Table \ref{tab:comparisons-to-ippo}, we show that in 4 out of 6 of the cooperative scenarios (Simple Spread, Joint Passage, Sampling, and Flocking), where coordinated actions among agents are necessitated on top of independent exploration, CoHet variants outperform IPPO by a large margin. Furthermore, the results show that, on average, CoHet outperforms HetGPPO by a factor of approximately 3.19.

\begin{table}[h!]
\centering
\caption{Mean Intrinsic Rewards of the two CoHet variants for an agent after 1000 iterations in the shared environment}
\label{tab:team-vs-self}
\begin{tabular}{@{}p{2.5cm}p{1.5cm}p{1.5cm}p{1.5cm}@{}} \toprule
\textbf{Scenario} & \textbf{CoHet\textsubscript{team}} & \textbf{CoHet\textsubscript{self}} & \textbf{Diff. $|\delta|$} \\ \midrule
Flocking & -0.0013 & -0.0023 & 0.0010 \\ 
Navigation & -0.0052 & -0.0027 & 0.0025 \\ 
Reverse Transport & -0.0011 & -0.0050 & 0.0039 \\ 
Sampling & -0.2289 & -1.118 & 0.8891 \\ 
Simple Spread & -0.3458 & -0.1890 & 0.1568 \\ 
Joint Passage & -0.0145 & -0.0249 & 0.0104 \\ \bottomrule
\end{tabular}
\end{table}
\subsection{CoHet\texorpdfstring{\textsubscript{team}}{ (team)} vs. CoHet\texorpdfstring{\textsubscript{self}}{ (self)}: A Comparative Analysis}
\label{sec:cohet-team-vs-cohet-self}
CoHet\textsubscript{team} utilizes the dynamics models of its surrounding agents to calculate intrinsic reward, whereas CoHet\textsubscript{self} employs the agent's own dynamics model and learns to follow its own predictions. As a result, the agents in CoHet\textsubscript{team} learn to behave in accordance with their neighborhood predictions, and this can be challenging. As the agent dynamics models are trained using their own interactions with the environment, the presence of heterogeneous agents with differing physical/behavioral capabilities can result in a diverse set of dynamics models. Subsequently, the agents failing to align with the predictions of their local neighborhood, consisting of diverse agents, will be heavily penalized. Results indicate that, unlike previous methods, the novel GNN-based formulation in CoHet allows the agents to more accurately model their local heterogeneous neighborhood. In Table \ref{tab:team-vs-self}, we demonstrate that the intrinsic reward penalty for misalignment with neighbors in CoHet\textsubscript{team} remains similar to the ones for CoHet\textsubscript{self}.\par
In case of the performance, CoHet\textsubscript{team} demonstrates better performance in all of the tasks that benefit from inter-agent cooperation, as it incentivizes the agents to adopt a more collaborative approach, irrespective of physical or behavioral heterogeneity. CoHet\textsubscript{self} exhibits superior performance over CoHet\textsubscript{team} in only the MPE Simple Spread task. Since the landmarks are not specifically assigned to specific agents in this task, we find that the agents can gain an advantage from exploiting known areas of the environment where any of the landmarks exist, thus they are encouraged to exploit the parts of the environments where the errors of their own dynamics models are minimal.
\begin{figure}[h]
    \centering
    \captionsetup[subfigure]{justification=centering}
    \begin{subfigure}{0.36\textwidth}
        \centering
        \includegraphics[width=.93\textwidth,height=0.175\textheight]{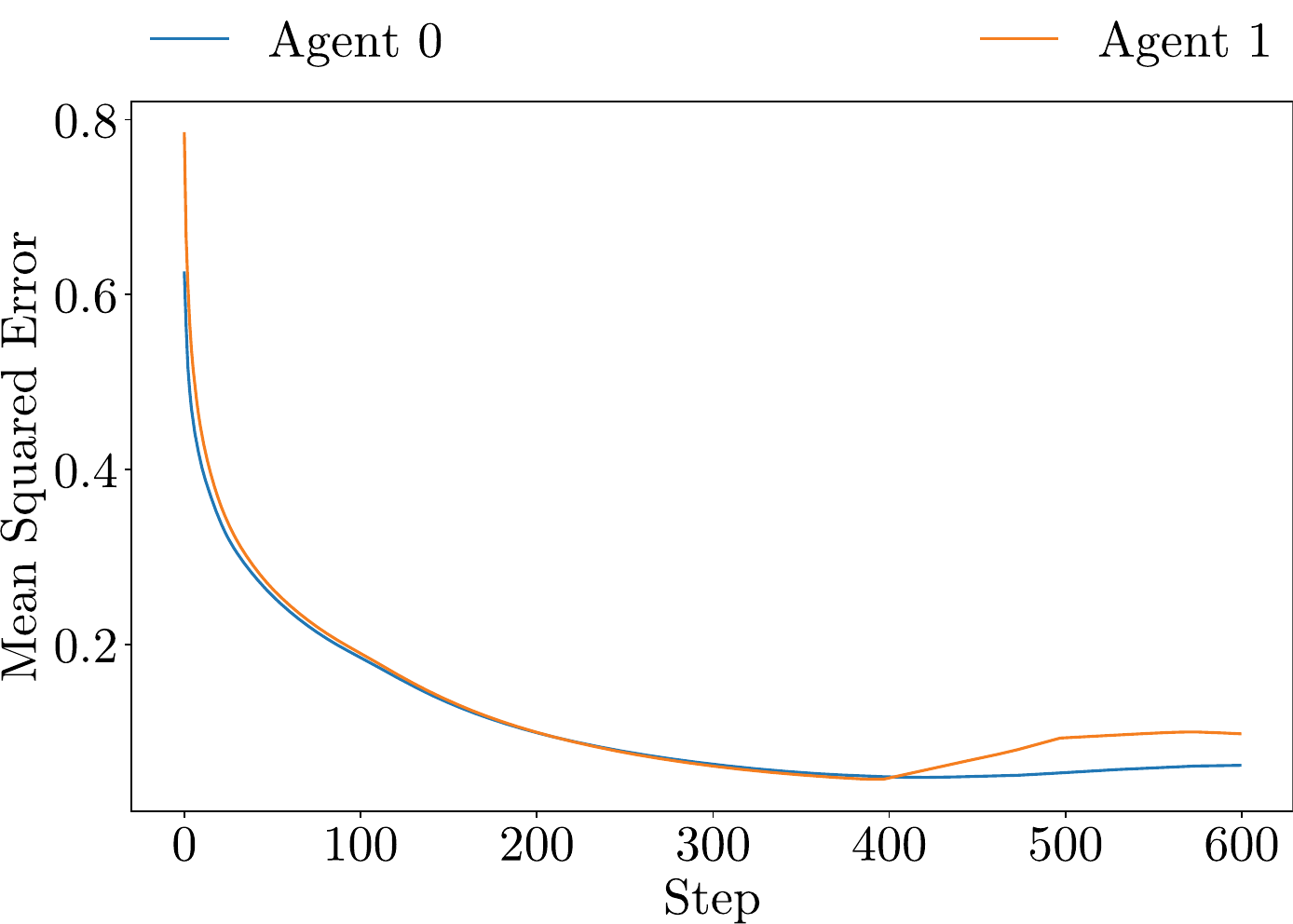}
        \caption{Mean Dynamics Model Loss}
        \label{subfig:dyn-model-loss}
    \end{subfigure}
    \hfill
    \begin{subfigure}{0.36\textwidth}
        \centering
        \includegraphics[width=.95\textwidth,height=0.18\textheight]{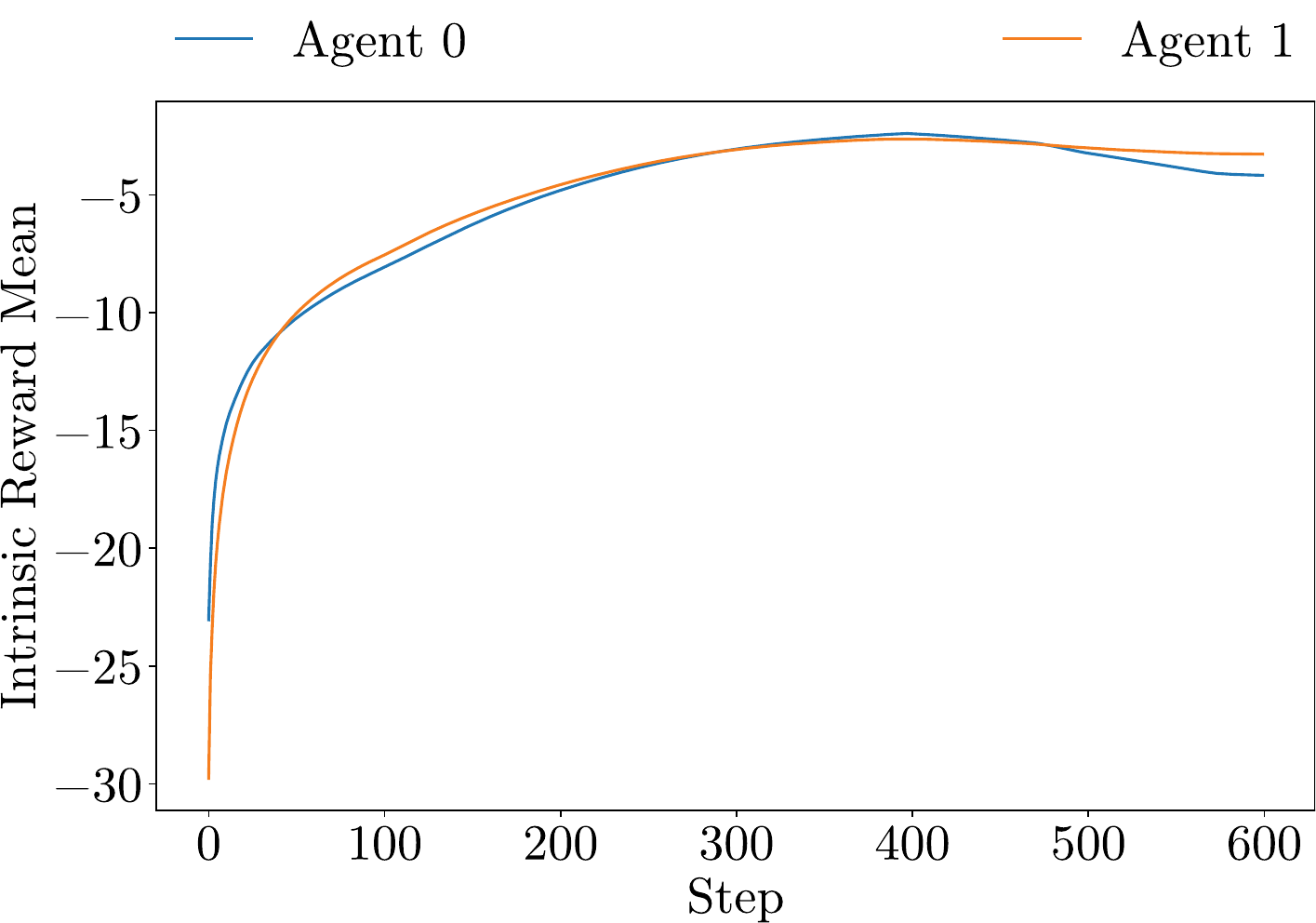}
        \caption{Mean Intrinsic Reward}
        \label{subfig:int-rew-decline}
    \end{subfigure}
    \caption{Reward architecture evaluation for two agents in the MPE Joint Passage task}
    \label{fig:int-reward-eval}
    \Description{Figure showing the reward architecture evaluation for two agents in the MPE Joint Passage task. The figure is divided into two subfigures: (a) Mean Dynamics Model Loss, which displays the loss metric of the dynamics model used for intrinsic reward calculation, and (b) Mean Intrinsic Reward, which shows the average intrinsic rewards received by the agents. Subfigure (a) illustrates how well the dynamics model predicts the agent's behavior, while subfigure (b) reflects the effectiveness of the intrinsic reward mechanism in guiding the agents. The evaluation is conducted over the MPE Joint Passage task, comparing the performance of the reward architecture.}
\end{figure}
\subsection{Reward Architecture Evaluation}
\label{sec:dyn-model-eval}
As the dynamics models for each agent continue to be trained, we anticipate that the agents will gain a better understanding of the environmental dynamics, leading to a decrease in the dynamics model Mean Squared Error (MSE) loss over time. Considering that each agent operates with their independent dynamics models, we should see a decline in the dynamics model loss for each. From our experiments, we observe this aforementioned gradual decline in the mean dynamics model loss for each agent, as depicted in Figure \ref{subfig:dyn-model-loss} for the MPE Joint Passage task. This decline is also observed in each task of Figure \ref{fig:mean-extrinsic-rewards}. As evident in Figure \ref{subfig:dyn-model-loss}, the agents effectively learn the environmental dynamics within 600 training episodes, and the dynamics model loss remains minimal unless an unknown aspect of the environmental dynamics is encountered. As a result of a better understanding of the environmental dynamics, each agent is more capable of predicting its neighbors’ next observations. Furthermore, the agents gradually adapt to meet their neighbors’ predictions. Therefore, the intrinsic reward, which in the case of CoHet is a penalty for misalignment, transitions from large negative values to extremely small negative values over time. This trend is demonstrated in Figure \ref{subfig:int-rew-decline} for the same task. Interestingly, as the intrinsic reward is calculated using the dynamics model loss, we can see that it exhibits the opposite trend to that of the loss.
\begin{table}[h!]
\centering
\caption{Robustness of CoHet\textsubscript{team} against increasing number of agents in VMAS-Navigation}
\label{tab:scalability}
\begin{tabular}{@{}c@{\hskip 1em}c@{\hskip 1em}c@{}} \toprule
\textbf{No. of agents} & \textbf{Mean Episodic Rewards} & \textbf{$\Delta$ vs 1 agent} \\ \midrule
1 agent & 0.9983 & 0.0000 \\ 
2 agents & 1.6803 & +0.6820 \\ 
3 agents & 1.7812 & +0.7829 \\ 
4 agents & 1.9698 & +0.9715 \\ 
8 agents & 2.6768 & +1.6785 \\ 
16 agents & 2.9670 & +1.9687 \\ \bottomrule
\end{tabular}
\end{table}
\subsection{Robustness to Increasing No. of Agents}
\label{sec:scalability}
As discussed previously, CoHet\textsubscript{team} motivates the agents to exhibit behaviors that match their neighborhood predictions. However, the diversity of heterogeneity types in the agent neighborhood can pose a challenge, potentially leading to a performance decline as the number of heterogeneous agents in the shared environment increases. We find that CoHet\textsubscript{team} maintains its robustness despite the growth in the number of heterogeneous agents in shared environments, indicated by the Mean Episodic Rewards in Table \ref{tab:scalability} after 1000 environmental episodes. The mean episodic rewards remain consistent with the rewards for the single-agent case, indicating the robustness of the proposed architecture to an increasing number of heterogeneous agents.
\section{Conclusions and Future Work}
\label{sec:conclusion}
In this paper, we introduce CoHet, a novel intrinsic reward mechanism leveraging a Graph Neural Network (GNN) for decentralized heterogeneous MARL policy learning. CoHet can be integrated with existing decentralized policy optimization methods and is well-suited for practical multi-agent systems facing partial observability and reward sparsity. We validate CoHet across cooperative multi-agent tasks involving heterogeneous agents, explore the impact of agent dynamics on the intrinsic reward module, and show that CoHet\textsubscript{team} remains robust as the number of heterogeneous agents increases.\par
A promising direction for future work includes exploring the alternative types of intrinsic motivation such as curiosity-driven or novelty-based rewards in the context of decentralized heterogeneous policy learning frameworks. Finding the proper balance between intrinsic and extrinsic rewards remains an open research problem. Future works can explore utilizing other types of weighting mechanisms that prioritize the predictions of the agents with the same sub-goals, and heterogeneity types. Ultimately, we opine that the collaboration between the agents in a MARL setting should take the need for decentralized training and agent heterogeneity into account.


\bibliographystyle{ACM-Reference-Format} 
\bibliography{AAMAS_2025_sample}


\end{document}